\documentclass[12pt]{article}

\usepackage{graphicx}

\begin{document}
\begin{center}

{\bf A new $F(R)$ gravity model}\\
\vspace{5mm}
 S. I. Kruglov
\footnote{serguei.krouglov@utoronto.ca}

\vspace{5mm}
\textit{Department of Chemical and Physical Sciences, University of Toronto,\\
3359 Mississauga Rd. North, Mississauga, Ontario, Canada L5L 1C6}
\end{center}

\begin{abstract}
We propose a new model of modified $F(R)$ gravity theory with the function $F(R) = (1/\beta) \arcsin(\beta R)$. Constant curvature solutions corresponding to the flat and de Sitter spacetime are obtained. The Jordan and Einstein frames are considered; the potential and the mass of the scalar degree of freedom are found. We show that the flat spacetime is stable and the de Sitter spacetime is unstable. The slow-roll parameters $\epsilon$, $\eta$, and the $e$-fold number of the model are evaluated in the Einstein frame. The index of the scalar spectrum power-law $n_s$ and the tensor-to-scalar ratio $r$ are calculated.
Critical points of autonomous equations for the de Sitter phase and the matter dominated epoch are found and studied. We obtain the approximate solution of equations of motion which is the deviation from the de Sitter phase in the Jordan frame. It is demonstrated that the model passes the matter stability test.
\end{abstract}


\section{Introduction}

It is possible to describe the inflation and the present time universe acceleration if one modifies the Einstein-Hilbert (EH) action of general relativity (GR). We propose the particular model of the $F(R)$ gravity with the help of replacing the Ricci scalar by the function $ F(R)=(1/\beta) \arcsin(\beta R)$ in EH action, where $\beta$ is the parameter with the dimension of (length)$^2$. Thus, we introduce the fundamental length $\sqrt{\beta}$
which goes probably from quantum gravity. The $F(R)$ gravity models may describe the evolution of the universe without introducing Dark Energy (DE) (Appleby, Battye and Starobinsky 2010, Capozziello and Faraoni 2011, Nojiri and Odintsov 2011). In such models the cosmic acceleration occurs due to modified gravity. Therefore, $F(R)$ gravity models can be an alternative to $\Lambda$-Cold Dark Matter ($\Lambda$CDM) model as new gravitational physics is considered. The $\Lambda$CDM model has a problem with the explanation of the smallness of the cosmological constant $\Lambda$. It should be mentioned that the form of the function $F(R)$ has to be derived from the fundamental theory (string, M-theory) which is absent. Therefore, different $F(R)$ gravity models that satisfies the general conditions are of interest.
The motivation for this work is to consider new F(R) model which meets requirements such as quantum and classical stabilities, it passes matter stability test and describes inflation of universe etc.

It should be noted that the first successful models of $F(R)$ gravity were given in Starobinsky 1980,
Hu and Sawicki 2007, Appleby and Battye 2007, Starobinsky 2007, Nojiri  and Odintsov 2007a, Nojiri and Odintsov 2008, Cognola et al 2008. Some $F(R)$ gravity models were introduced in Deser and Gibbons 1998, Capozziello and Faraoni 2011, Kruglov 2013, Kruglov 2014a, Kruglov 2014b and in other publications.
$F(R)$ gravity models are phenomenological models that may describe different eras and the evolution of the universe. The first $F(R)$ gravity model was introduced in Starobinsky 1980 that gives the self-consistent description of the inflation.

The paper is organized as follows. We formulate the model with one dimensional parameter $\beta$ in section 2. It is shown that the classical and quantum stabilities take place in the model under consideration. We obtain the constant curvature solutions corresponding to flat spacetime, $R_0=0$, and to the Schwarzschild-de Sitter spacetime, $\beta R_0\approx 0.919$. In section 3 the scalar-tensor formulation of the model is investigated (in the Einstein frame). The potential and the mass of the scalar degree of freedom (scalaron) are found. We obtain the slow-roll parameters $\epsilon$, $\eta$, and the $e$-fold number of the model in section 4. The index of the scalar spectrum power-law $n_s$ and the tensor-to-scalar ratio $r$ are calculated. Critical points of autonomous equations for the de Sitter phase and the matter dominated epoch are found in section 5. The approximate solution of equations of motion in the Jordan frame corresponding to the deviation from the de Sitter phase is obtained in section 6. We show in section 7 that the model passes the matter stability test. Section 8 is devoted to a conclusion.

The Minkowski metric $\eta_{\mu\nu}$=diag(-1, 1, 1, 1) is used and $c$=$\hbar$=1 is assumed.

\section{The Model}

Let us consider a new model of arcsin-gravity with the Lagrangian density
\begin{equation}
{\cal L} =\frac{1}{2\kappa^2}F(R) =\frac{1}{2\kappa^2}\left[\frac{1}{\beta}\arcsin(\beta R)\right],
\label{1}
\end{equation}
where $\kappa=M_{Pl}^{-1}$, $M_{Pl}$ is the reduced Planck mass, $\beta$ has the dimension of (length)$^2$, and the action without matter is given by $S=\int d^4x\sqrt{-g}{\cal L}$.
At $\beta R\ll 1$, we have $\arcsin(\beta R)\approx \beta R$, and  we arrive at the EH action. The equation $F(0)=0$ holds, corresponding to the flat space-time without cosmological constant. GR passes local tests and we imply that at the present time the low curvature regime occurs, $\beta R\ll 1$. We will describe the inflation and universe evolution in the model suggested.
For the classical stability the inequality $F'(R)>0$ (the prime means the derivative with the respect to the argument) is required (Appleby, Battye and Starobinsky 2010) which is satisfied if $\beta R<1$,
\begin{equation}
F'(R)=\frac{1}{\sqrt{1-(\beta R)^2}}>0.
\label{2}
\end{equation}
Quantum stability claims the inequality $F''(R)> 0$ (Appleby, Battye and Starobinsky 2010), that becomes in our model as follows:
\begin{equation}
F''(R)=\frac{\beta^2 R}{\left[1-(\beta R)^2\right]^{3/2}}>0,
\label{3}
\end{equation}
and it is also satisfied at $0<\beta R<1$.

\subsection{Constant Curvature Solutions}

If the Ricci scalar $R$ is a constant, $R=R_0$, equations of motion (Barrow, 1983) become
\begin{equation}
2F(R_0)=R_0F'(R_0),
\label{4}
\end{equation}
and are given, in the model with the Lagrangian density (1), by
\begin{equation}
2\sqrt{1-(\beta R_0)^2}\arcsin (\beta R_0)=\beta R_0.
\label{5}
\end{equation}
We note that constant curvature solutions correspond to the extremum of the effective potential.
It should be mentioned that the general conditions for multiply de Sitter solutions in $F(R)$ gravity were discussed in Cognola et al 2005, and Cognola et al 2009. The fact of the appearance of such solutions is known for F(R) gravity theories with many parameters. But the model (1) has only one parameter $\beta$ that is, in our opinion, an attractive feature of the model. Therefore, it is of interest to analyze one more $F(R)$ gravity model.

Eq. (5) possesses two solution, $R_0=0$ corresponding to the flat spacetime, and non-trivial solution $\beta R_0\approx 0.919$. We will show that the last solution goes with the Schwarzschild-de Sitter spacetime and with the maximum of the effective potential in the Einstein frame.
The constant curvature solutions describe the acceleration phase which is future stable if the inequality $F'(R_0)/F''(R_0)>R_0$ occurs (M\"{u}ller, Schmidt and Starobinsky 1988), and we have from Eqs. (2),(3)
\begin{equation}
1-\left(\beta R\right)^2>\left(\beta R\right)^2,
\label{6}
\end{equation}
which is equivalent to $\beta R<1/\sqrt{2}\approx 0.707$. Thus, the solution $R_0=0$ obeys Eq. (6) and the flat spacetime is stable. The second constant curvature solution $\beta R_0\approx 0.919$ does not satisfy Eq. (6), leads to unstable de Sitter spacetime, and describes the inflation.

\section{The Scalar-Tensor Formulation}

\begin{figure}[h]
\includegraphics[height=4.0in,width=4.0in]{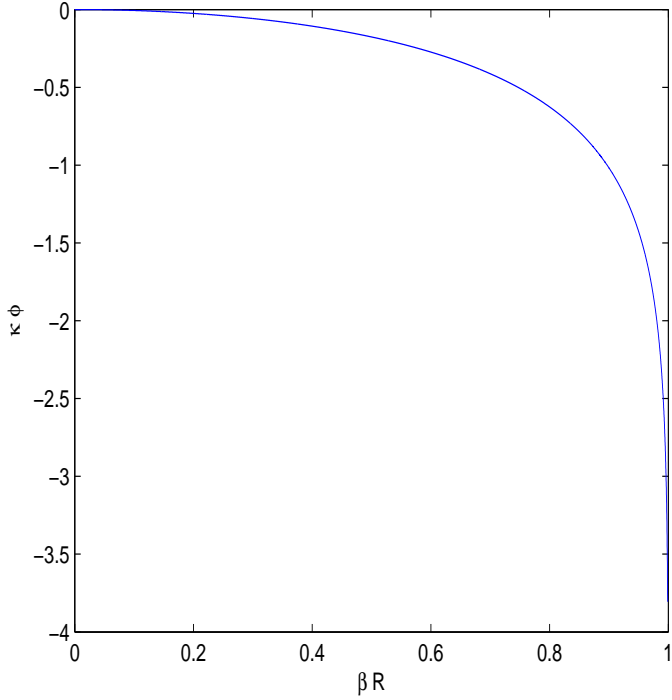}
\caption{\label{fig.1}The function  $\kappa\phi$ versus $\beta R$. }
\end{figure}

\begin{figure}[h]
\includegraphics[height=4.0in,width=4.0in]{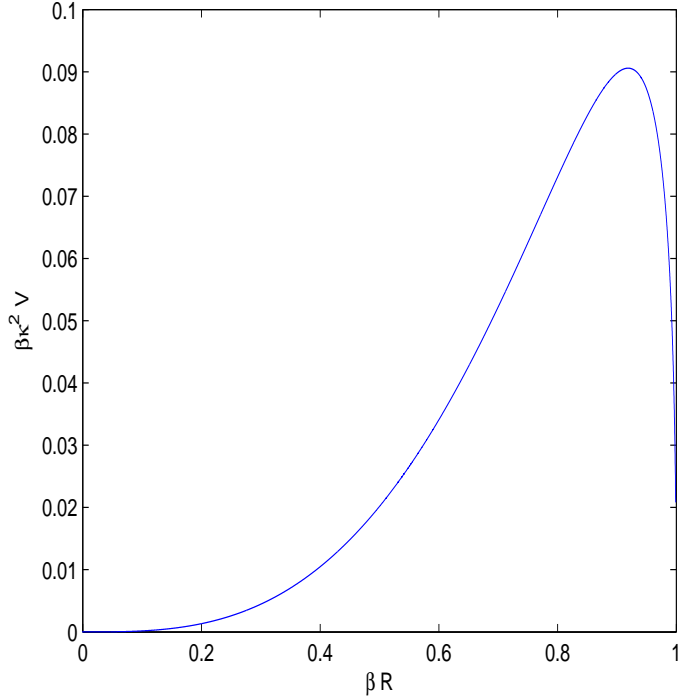}
\caption{\label{fig.2}The function $\beta\kappa^2V$ versus $\beta R$.}
\end{figure}

\begin{figure}[h]
\includegraphics[height=4.0in,width=4.0in]{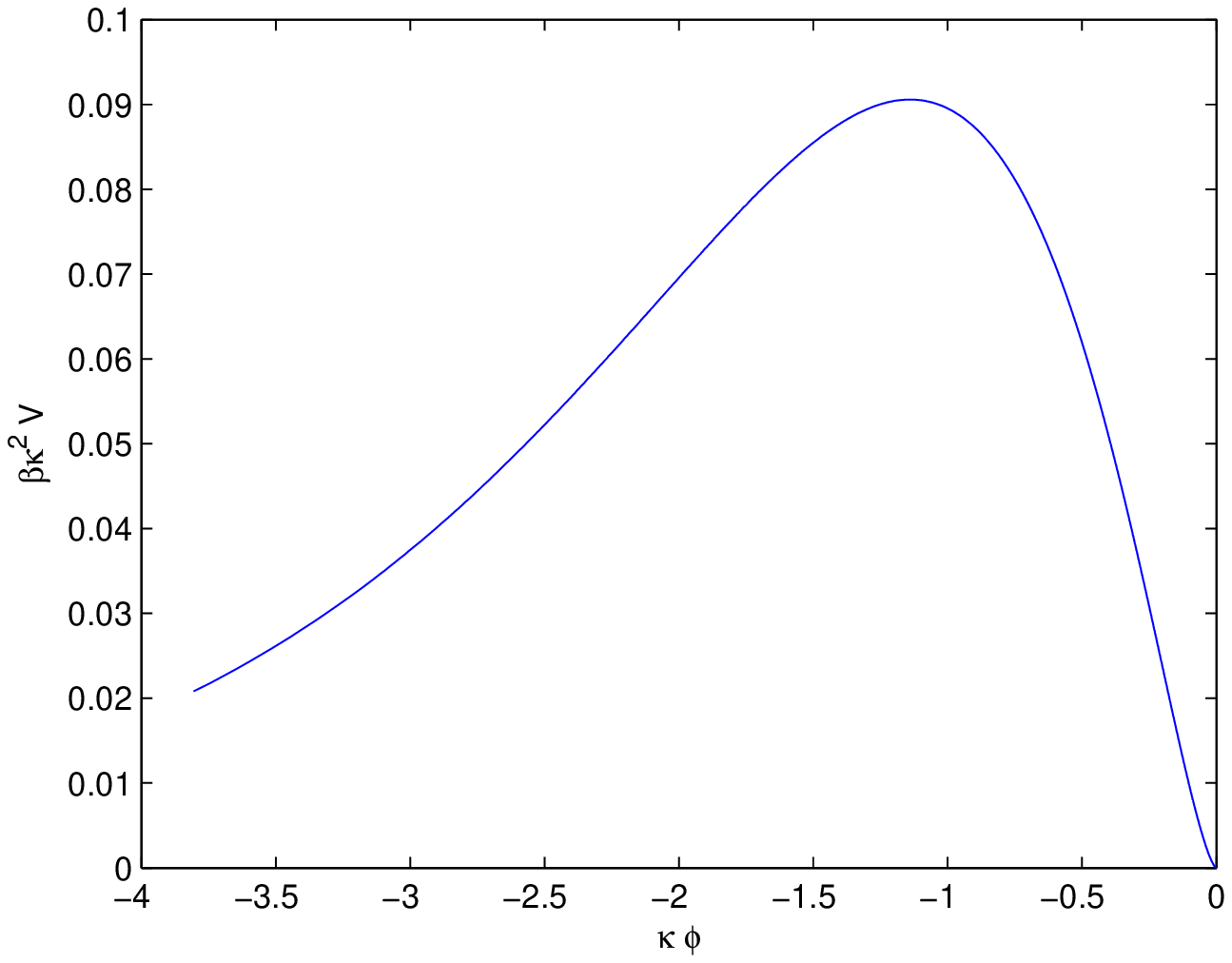}
\caption{\label{fig.3}The function $\beta\kappa^2V$ versus $\kappa\phi$.}
\end{figure}

Now we investigate the model in Einstein's frame performing the conformal transformation of the metric (Magnano and Sokolowski 1994)
\begin{equation}
\widetilde{g}_{\mu\nu} =F'(R)g_{\mu\nu}=\frac{1}{\sqrt{1-(\beta R)^2}}g_{\mu\nu}.
\label{7}
\end{equation}
Then the Lagrangian density in Einstein's frame becomes
\begin{equation}
{\cal L}=\frac{\widetilde{R}}{2\kappa^2}-\frac{1}{2}\widetilde{g}^{\mu\nu}
\nabla_\mu\phi\nabla_\nu\phi-V(\phi).
\label{8}
\end{equation}
Here the Ricci scalar $\widetilde{R}$ in the Einstein frame is calculated in new metric (7) and the scalar field $\phi$ is \begin{equation}
\phi(R)=-\frac{\sqrt{3}}{\sqrt{2}\kappa}\ln F'(R)=\frac{\sqrt{3}}{\sqrt{2}\kappa}\ln\sqrt{1-(\beta R)^2}.
\label{9}
\end{equation}
The plot of the function $\kappa\phi$ is presented in Fig. \ref{fig.1}.
The potential $V(\phi)$ is given by
\[
V(R)=\frac{RF'(R)-F(R)}{2\kappa^2F'^2(R)}
\]
\vspace{-7mm}
\begin{equation}
\label{10}
\end{equation}
\vspace{-7mm}
\[
=\frac{1}{2\beta \kappa^2}\left[\beta R\sqrt{1-(\beta R)^2}-\left(1-(\beta R)^2\right)\arcsin(\beta R)\right].
\]
The plot of the function $\beta\kappa^2V$ versus $\beta R$ is given in Fig. \ref{fig.2} and the plot of function $\beta\kappa^2V$ versus $\kappa\phi$ is represented by Fig. \ref{fig.3}.
The extremum of the potential, $V'(R)=0$, with the help of Eq. (10) leads to Eq. (4).
The potential (10) possesses the minimum at $R=0$ and the maximum at $\beta R_0\approx 0.919$. The flat space-time ($R=0$) is the stable state and the state with the curvature $R_0\approx 0.919/\beta$ is unstable.

\begin{figure}[h]
\includegraphics[height=4.0in,width=4.0in]{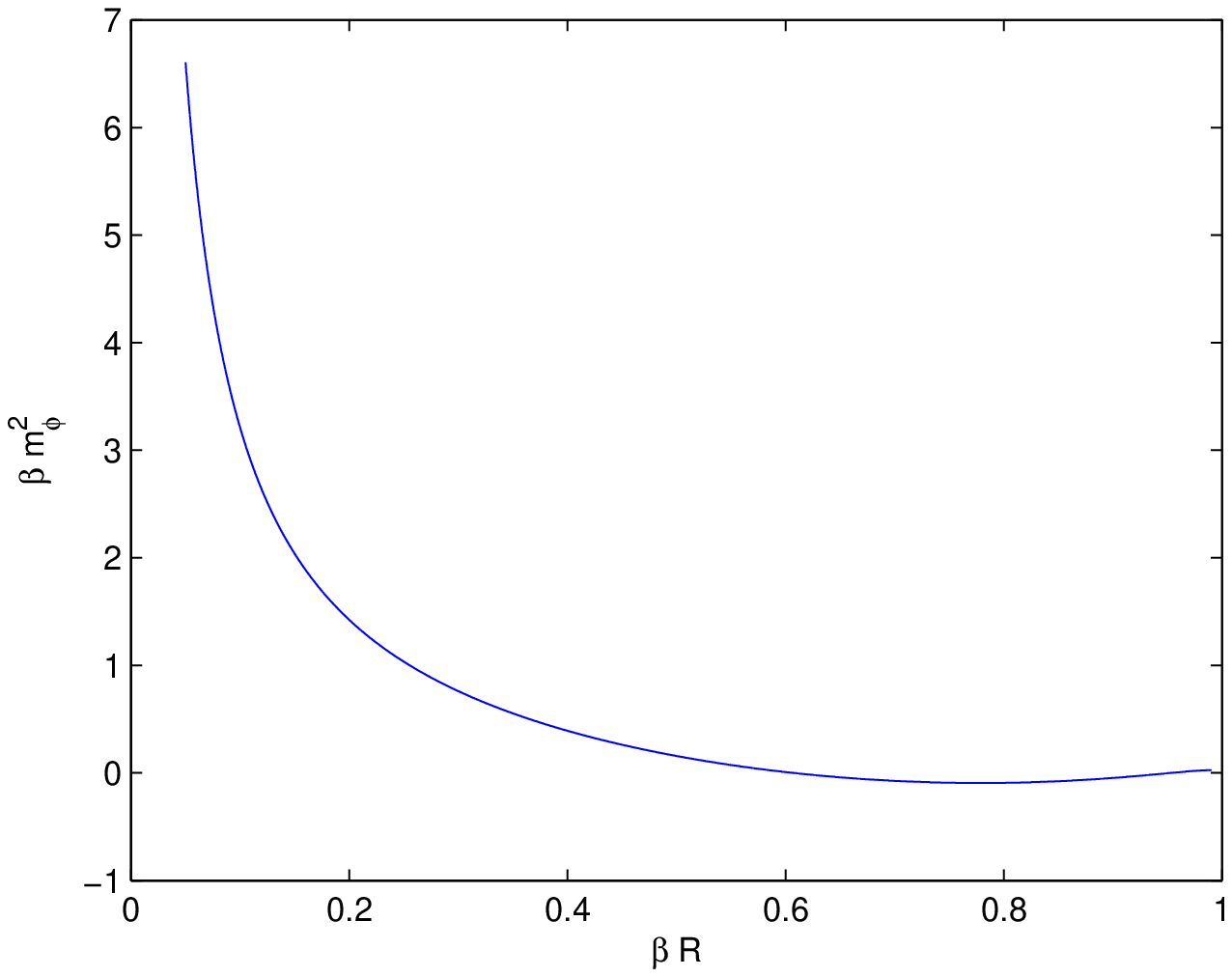}
\caption{\label{fig.4}The function $\beta m^2_\phi$ versus $\beta R$. }
\end{figure}

We obtain the mass squared of a scalaron (scalar degree of freedom) from Eq. (10),
\[
m_\phi^2=\frac{d^2V}{d\phi^2} =\frac{1}{3}\left(\frac{1}{F''(R)}+\frac{R}{F'(R)}-\frac{4F(R)}{F^{'2}(R)}\right)
\]
\begin{equation}
=\frac{1}{3\beta}\biggl[\frac{\left(1-x^2\right)^{3/2}}
{x}+x\sqrt{1- x^2}-4\left(1- x^2\right)\arcsin x\biggr],
\label{11}
\end{equation}
where $x=\beta R$. The plot of the function $\beta m_\phi^2$ versus $x=\beta R$ is given by Fig. \ref{fig.4}.
One can verify that $m^2_\phi<0$ for the constant curvature solution $R_0\approx 0.919/\beta$, and, therefore, this solution corresponds to unstable state as it was mentioned before. It follows from Eq. (11) that at $0.529<\beta R<1$ we have non-stable states, $m^2_\phi<0$. The stability of the de Sitter solution in F(R) gravity models was first studied by M\"{u}ller, Schmidt and Starobinsky 1988. To pass the Solar system tests the value $m^2_\phi$ should be positive and big.
If the value $\beta R$ is small the mass $m_\phi$ is big according to Fig. \ref{fig.4} and corrections to the Newton law are negligible.

To assure that corrections of $F(R)$ gravity model are small as compared to GR for $R\gg R_1$, where $R_1$ is a curvature at the present time, the relations
\begin{equation}
\mid F(R)-R\mid < R,~~\mid F'(R)-1\mid < 1,~~\mid RF''(R)\mid< 1
\label{12}
\end{equation}
should hold (Appleby and Battye 2010).
As $\arcsin x>x$ at $1>x>0$ ($x=\beta R$), the first inequality in Eq. (12) becomes $\arcsin x< 2x$, and it is satisfied at $1>x>0$. The second inequality in Eq. (12) is equivalent to $x<\sqrt{3}/2\approx 0.866$ (as $F'(R)>1$ for $0<x<1$). The third inequality in Eq. (12) holds at $0<x<0.655$.
As a result, all Eqs. (12) are satisfied if $0<x<0.655$.

\section{Slow-Roll Cosmological Parameters}

\begin{figure}[h]
\includegraphics[height=4.0in,width=4.0in]{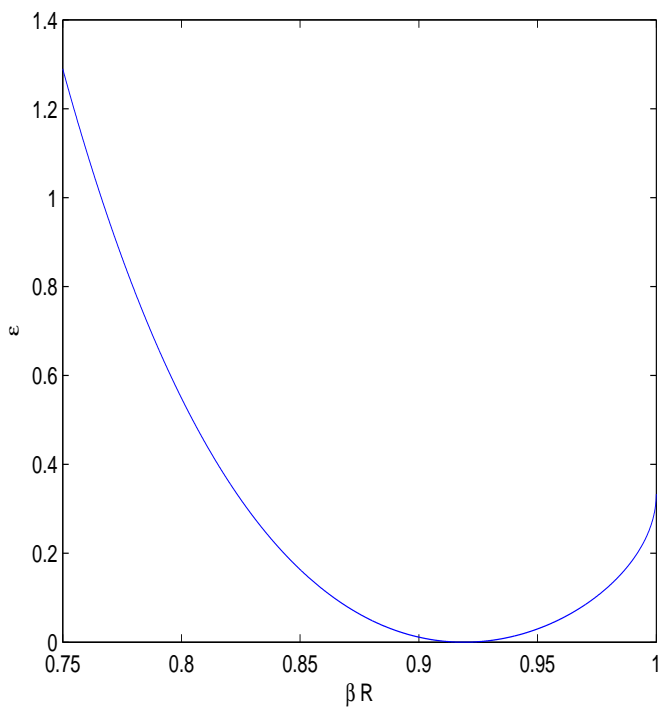}
\caption{\label{fig.5} The function $\epsilon$ versus $\beta R$.}
\end{figure}

\begin{figure}[h]
\includegraphics[height=4.0in,width=4.0in]{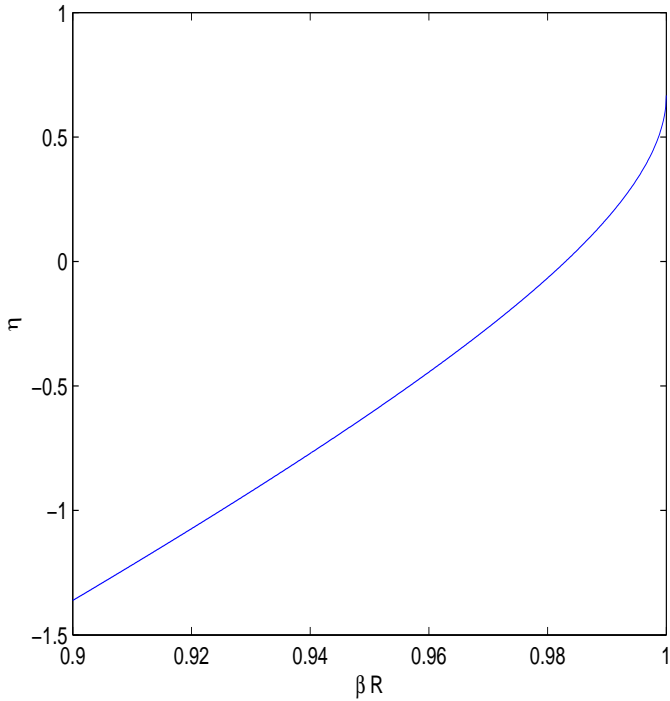}
\caption{\label{fig.6}The function $\eta$ versus $\beta R$. }
\end{figure}

The slow-roll parameters are given by (Liddle and Lyth 2000)
\begin{equation}
\epsilon(\phi)=\frac{1}{2}M_{Pl}^2\left(\frac{V'(\phi)}{V(\phi)}\right)^2,~~~~\eta(\phi)=M_{Pl}^2\frac{V''(\phi)}{V(\phi)}.
\label{13}
\end{equation}
When conditions $|\eta(\phi)|\ll 1$, $\epsilon(\phi)\ll 1$ hold the slow-roll approximation takes place.
From Eqs. (10),(11) we find the slow-roll parameters as follows:
\[
\epsilon=\frac{1}{3}\left[\frac{RF'(R)-2F(R)}{RF'(R)-F(R)}\right]^2
\]
\begin{equation}
=\frac{1}{3}\left(\frac{2\sqrt{1-x^2}\arcsin x-x}{\sqrt{1-x^2}\arcsin x-x}\right)^2,
\label{14}
\end{equation}
\[
\eta=\frac{2}{3}\left[\frac{F^{'2}(R)+F''(R)\left[RF'(R)-4F(R)\right]}{F''(R)\left[RF'(R)-F(R)\right]}\right]
\]
\begin{equation}
=\frac{2\left(1-4x\sqrt{1-x^2}\arcsin x\right)}{3x\left(x-\sqrt{1-x^2}\arcsin x\right)}.
\label{15}
\end{equation}
The plots of the functions $\epsilon$, $\eta$ are given in Fig. \ref{fig.5}, Fig. \ref{fig.6}.
The equation $\epsilon =1$ has the solution $x\approx 0.766$. It follows from Fig. \ref{fig.5} that at $1>\beta R>0.766$ the inequality $\epsilon <1$ holds. The equation $|\eta | =1$ is satisfied at $x\approx 0.516$, $x\approx 0.544$ and $x\approx 0.925$. At $0.544>\beta R>0.516$ and at $1>\beta R>0.925$, we have the result $|\eta | < 1$. Therefore, the slow-roll approximation, $\epsilon <1$ and $|\eta | < 1$, takes place at $1>\beta R>0.925$.

The age of the inflation can be obtained by calculating the $e$-fold number (Liddle and Lyth 2000)
\begin{equation}
N_e\approx \frac{1}{M_{Pl}^2}\int_{\phi_{end}}^{\phi}\frac{V(\phi)}{V'(\phi)}d\phi.
\label{16}
\end{equation}
We find, from Eqs. (9),(10), the number of $e$-foldings
\begin{equation}
N_e\approx \frac{3}{2}\int_{x_{end}}^{x_0}\frac{x\left(\sqrt{1-x^2}\arcsin x-x\right)dx}
{\left(1-x^2\right)\left(2\sqrt{1-x^2}\arcsin x-x\right)},
\label{17}
\end{equation}
were $x_{end}=\beta R_{end}$ corresponds to the time of the end of inflation when $\epsilon$ or $|\eta|$ are close to $1$. Thus, inflation ends when slow-roll conditions are violated. We obtain the amount of inflation $N_e\approx 9.7$ at $x_0=0.9999$ and $x_{end}=0.92$, and, therefore, the model can describe the inflation. It should be noted that it is required around $60$ $e$-foldings of inflation to solve the flatness and horizon problems.

Due to density perturbations the index of the scalar spectrum power-law is given by the relation (Liddle and Lyth 2000)
\begin{equation}
n_s=1-6\epsilon+2\eta.
\label{18}
\end{equation}
Using Eqs. (14),(15), the plot of the function of $n_s$ versus $\beta R$ is represented in the Fig.7.
\begin{figure}[h]
\includegraphics[height=4.0in,width=4.0in]{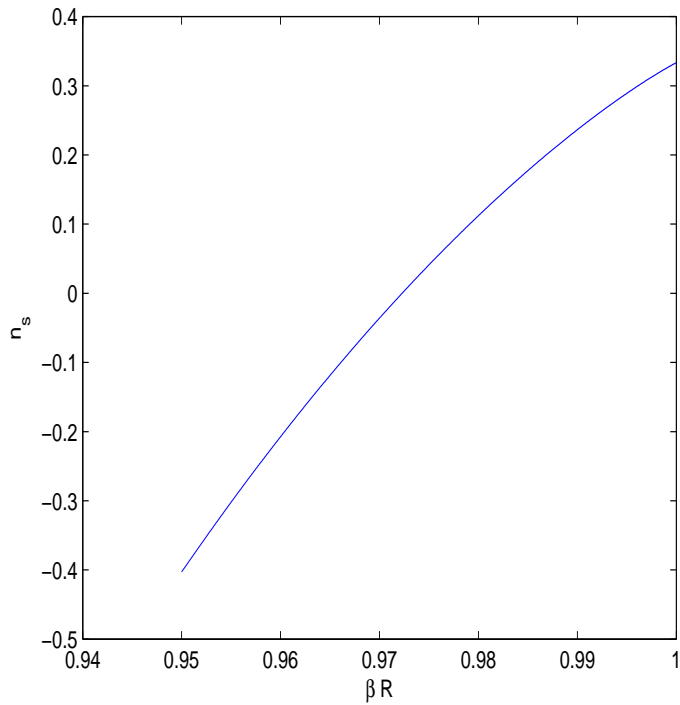}
\caption{\label{fig.7}The function $n_s$ versus $\beta R$.}
\end{figure}
The tensor-to-scalar ratio is defined by Liddle and Lyth 2000, $r=16\epsilon$. The PLANCK experiment gives the result (Ade et al 2014)
\begin{equation}
n_s=0.9603\pm 0.0073,~~~~r<0.11.
\label{19}
\end{equation}
One can see from Fig. 7 that the experimental value of $n_s$ is not satisfied. But the bound for tensor-to-scalar ratio $r<0.11$ is satisfied for $0.933>\beta R>0.905$. As a result, the model suggested can give only approximate description of cosmology in the Einstein frame.

It was stated in Bamba at al 2014, Bamba and Odintsov 2015 that cosmology in the Einstein and Jordan frames can be different and these frames are physically non-equivalent. As a result, theories in the Einstein and Jordan frames may be considered as different cosmological theories. One can recalculate inflationary parameters in the $F(R)$ frame (see Bamba at al 2014).

\section{Critical Points of Autonomous Equations}

To investigate critical points of equations of motion in the Jordan frame, it is useful to introduce the dimensionless parameters (Amendola et al 2007) which become
\[
x_1=-\frac{\dot{F}'(R)}{HF'(R)}=-\frac{x\dot{x}}{H\left(1-x^2\right)},
\]
\[
~x_2=-\frac{F(R)}{6F'(R)H^2}=-\frac{\sqrt{1-x^2}\arcsin x}{6\beta H^2},
\]
\begin{equation}
x_3=\frac{\dot{H}}{H^2}+2,
\label{20}
\end{equation}
\[
m=\frac{RF''(R)}{F'(R)}=\frac{x^2}{\left(1-x^2\right)},
\]
\begin{equation}
r=-\frac{RF'(R)}{F(R)}=\frac{x_3}{x_2}=-\frac{x}{\sqrt{1-x^2}\arcsin x},
\label{21}
\end{equation}
where $H$ is a Hubble parameter, $x=\beta R$, and the dot over the variables means the derivative with respect to the time. The deceleration parameter $q$ is given by $q=1-x_3$. Equations of motion in the absence of the radiation, $\rho_{rad}=0$, with the help of Eqs. (20), (21) can be written in the form of autonomous equations (Amendola et al 2007). One can investigate the critical points of the system of equations  by the study of the function $m(r)$ which shows the deviation from the $\Lambda$CDM model. The plot of the function $m(r)$ is presented by Fig. \ref{fig.7}.
\begin{figure}[h]
\includegraphics[height=4.0in,width=4.0in]{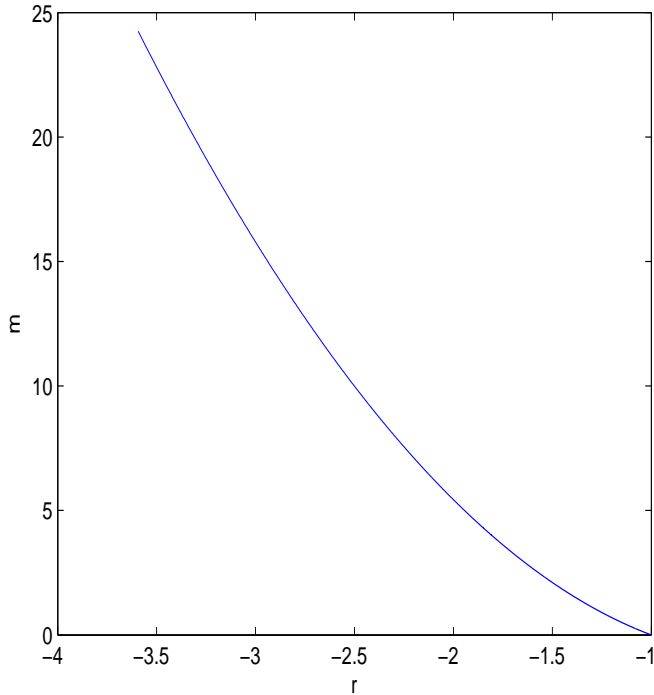}
\caption{\label{fig.7} The function $m(r)$}
\end{figure}
The de Sitter point $P_1$ (Amendola et al 2007), in the absence of radiation, $x_4 = 0$, corresponds to the parameters $x_1=0$, $x_2=-1$, $x_3=2$ ($\dot{H}=0$, $H^2=R/12$, $r=-2$). The point $P_1$ corresponds to the constant curvature solutions that may be verified using Eqs. (5),(20). The effective equation of state (EoS) parameter, $w_{eff}$, and the parameter of matter energy fraction, $\Omega_{m}$, are given for this point by
\begin{equation}
w_{eff}=-1-2\dot{H}/(3H^2)=-1,~~~~\Omega_{m}=1-x_1-x_2-x_3=0,
\label{22}
\end{equation}
which correspond to de Sitter phase. This point mimics a cosmological constant and the deceleration parameter becomes $q=-1$. The constant curvature solution $x \approx 0.919$ corresponds to unstable de Sitter space as $1< m(r=-2)\approx 5.4$ (Amendola et al 2007).

For the critical point $P_5$ ($x_3=1/2$), $m\approx 0$, $r\approx -1$, and EoS of a matter era is $w_{eff}=0$ ($a=a_0t^{2/3}$). Then we have a viable matter dominated epoch prior to the late-time acceleration (Amendola et al 2007). The equation $m=-r-1$ has the solution $m=0$, $r=-1$, $R=0$, corresponding to the point $P_5$.
One can verify with the help of Eq. (21) (see Fig. \ref{fig.7}) that $m'(r=-1)=0$. As a result, the condition $m'(r=-1)>-1$ holds and we have the standard matter era (Amendola et al 2007). Therefore, the correct description of the standard matter era occurs in the model under consideration. To investigate the possibility of late-time acceleration (DE) in the model, one needs to solve and analyze autonomous equations.
The unification of inflation with DE has been proposed first in $F(R)$ gravity by Nojiri and Odintsov 2003, and Nojiri  and Odintsov 2007b.

\section{Equations of Motion in the Jordan Frame and Their Approximate Solutions}

It should be mentioned that the change of the frame includes the change of the time scale and, therefore, the values of the slow-roll parameters are different in the frames. Now we consider equations of motion in the Jordan frame and their approximate solutions. If one adds to (1) the Lagrangian
of the matter with the energy-momentum tensor $T^{(m)}_{\mu\nu}$, we obtain equations of motion
\begin{equation}
R_{\mu\nu}F'(R)-\frac{1}{2}g_{\mu\nu}F(R)+g_{\mu\nu}g^{\alpha\beta}\nabla_\alpha\nabla_\beta F'(R) -\nabla_\mu\nabla_\nu F'(R)=\kappa^2T^{(m)}_{\mu\nu},
\label{23}
\end{equation}
where $\nabla_\mu$ is a covariant derivative. We consider the homogeneous, isotropic and spatially flat Friedmann-Robertson-Walker (FRW) cosmology with the line element
\begin{equation}
ds^2 = -dt^2 + a^2(t)\left(dx^2+dy^2+dz^2\right).
\label{24}
\end{equation}
Taking the trace of the left and right sides of Eq. (23) one finds
\begin{equation}
RF'(R)-2F(R)+3g^{\alpha\beta}\nabla_\alpha\nabla_\beta F'(R) =\kappa^2{\cal T},
\label{25}
\end{equation}
where ${\cal T}=g^{\mu\nu}T^{(m)}_{\mu\nu}$. For the FRW metric (24) we obtain the expression
\begin{equation}
g^{\alpha\beta}\nabla_\alpha\nabla_\beta R=\left(-g\right)^{-1/2}\partial_\mu\left[\left(-g\right)^{-1/2}g^{\mu\nu}\partial_\nu R\right]=-\ddot{R}-3H\dot{R}.
\label{26}
\end{equation}
Then with the help of Eqs. (1), (26) one finds the equation for the scalar curvature
\begin{equation}
\frac{R}{\sqrt{1-(\beta R)^2}}-\frac{2}{\beta}\arcsin (\beta R)-3\left(\frac{d^2}{dt^2}+3H\frac{d}{dt}\right)\frac{1}{\sqrt{1-(\beta R)^2}}=\kappa^2{\cal T}.
\label{27}
\end{equation}
The Ricci scalar can be expressed through the Hubble parameter as follows:
\begin{equation}
R = 12H^2 + 6\dot{H}.
\label{28}
\end{equation}
We consider the case ${\cal T}=0$, corresponding to the absence of the matter or the traceless of the electromagnetic fields. For the case of the constant curvature Eq. (27) converts to Eq. (4). There exists the exact solution to Eqs. (27), (28)
\begin{equation}
R_0 = 0,~~~H=\frac{1}{2t},~~~a(t)=a_0\sqrt{t},
\label{29}
\end{equation}
which is the same as the solution in GR. The solution (29) with the flat spacetime corresponds to the radiation era.
To describe all eras one should solve the system of nonlinear equations (27), (28). We consider the approximate solution to this system of equations which is the deviation from the constant curvature solution $\beta R_0\approx 0.919$. Thus, we expand $R=R_0+R_1$ that is the deviation from de Sitter phase, and assuming $R_1\ll R_0$. Linearizing Eq. (27), we obtain
\begin{equation}
\left[2(\beta R_0)^2-1\right]R_1 -3\beta^2 R_0\left(\ddot{R_1}+3H_0\dot{R_1}\right)= 0,
\label{30}
\end{equation}
where $H_0=\sqrt{R_0/12}$. Solutions to the linear Eq. (30) is in the form $R_1=A\exp (n_\pm t)$ with
\begin{equation}
n_\pm=\frac{-9H_0\beta R_0\pm \sqrt{81H_0^2(\beta R_0)^2+12R_0[2(\beta R_0)^2-1]}}{6\beta R_0}.
\label{31}
\end{equation}
The physical solution with the decreasing curvature is $R_1=A\exp (n_{-}t)$ which for $\beta R_0\approx 0.919$ becomes
\begin{equation}
R_1=A\exp\left(-\frac{1.066 t}{\sqrt{\beta}}\right).
\label{32}
\end{equation}
The linear equation (30) does not fix the amplitude $A$ that is small compared to $R_0$ so that $\beta A\ll 0.919$. Eq. (32) shows the rate of decreasing the curvature in the de Sitter point. Thus, the approximate solution to Eq. (27) at ${\cal T}=0$ for small deviation from the de Sitter phase is $\beta R=0.919+\beta A \exp\left(-1.066 t/\sqrt{\beta}\right)$. Thus, we propose $F(R)$ gravity model describing universe inflation and at weak curvature it becomes GR. The model under consideration is eternal F(R) inflation model. Some models of such sort were discussed by Nojiri and Odintsov 2015.
To describe all phases of the universe evolution one needs to solve the system of equations (27), (28) analytically or numerically. We leave this for further investigations.

\section{Matter Stability}

We follow the method of Dolgov and Kawasaki 2003 to investigate the matter stability. For weak gravity objects the Minkowski metric (flat) can be used and the approximate relation $g^{\alpha\beta}\nabla_\alpha\nabla_\beta\simeq \partial_k^2-\partial_t^2$ holds.
When $R$ is uniform (for spatially constant distribution) equation (25) becomes
\begin{equation}
-3F^{(2)}(R)\ddot{R}-3F^{(3)}(R)\dot{R}^2+ F^{(1)}(R)R-2F(R)=\kappa^2{\cal T},
\label{33}
\end{equation}
where $F^{(n)}(R)=d^nF(R)/dR^n$.
Let us consider a perturbative solution $R=R_0+R_1$ with $R_1$ being the perturbed part, $|R_1|\ll|R_0|$). According to GR the curvature in the lowest order is given by $R_0 =- \kappa^2{\cal T}$ inside the matter and $R_0 = 0$ outside the matter. Following to Nojiri  and Odintsov 2003, 2011, from equation (33), one finds
\begin{equation}
\ddot{R}_0+\ddot{R}_1+\frac{F^{(3)}(R_0)}{F^{(2)}(R_0)}\left(\dot{R}_0^2+2\dot{R}_0\dot{R}_1\right)+ \frac{2F(R_0)-R_0\left[1+F^{(1)}(R_0)\right]}{3F^{(2)}(R_0)}=U(R_0)R_1,
\label{34}
\end{equation}
where
\[
U(R_0)=\frac{F^{(3)2}-F^{(2)}F^{(4)}}{F^{(2)2}}\dot{R_0}^2
\]
\begin{equation}
+ \frac{\left(R_0F^{(2)}-F^{(1)}\right)F^{(2)}+\left(2F-R_0F^{(1)}-R_0\right)
F^{(3)}}{3F^{(2)2}}.
\label{35}
\end{equation}
The matter is unstable if $U(R_0)>0$ as $R_1$ exponentially increases in the time. From Eq. (1) we obtain the function $U(R_0)$:
\[
U(R_0)=\frac{1-5x^2-2x^4}{x^2(1-x^2)^2}\left(\beta\dot{R_0}\right)^2
+ V(R_0),
\]
\begin{equation}
V(R_0)=\frac{1}{3\beta}\left[\left(2\arcsin(x)-\frac{x}{\sqrt{1-x^2}}-x\right)\frac{(1+2x^2)\sqrt{1-x^2}}{x^2}
+\frac{2x^2-1}{x}\right],
\label{36}
\end{equation}
where $x=\beta R_0$.
One can check that $V(R_0)<0$ for $0<x<1$. Thus, for almost constant curvature, $\dot{R_0}\approx 0$, there is no matter instability, $U(R_0)<0$. As a result, this indicates on stability of the gravitational system and the model passes the matter stability test.

\section{Conclusion}

We propose a new model of modified $F(R)$ gravity representing the effective gravity model which can describe the evolution of the universe. The constant curvature solutions, $\beta R_0=0$, $\beta R_0=0.919$, were obtained that correspond to the flat spacetime and the de Sitter spacetime, correspondingly.
The de Sitter spacetime gives the acceleration of the universe and corresponds to the inflation. The flat space-time is stable but the de Sitter spacetime is unstable in the model and it goes with the maximum of the effective potential in the Einstein frame. The Jordan and Einstein frames were considered and we have obtained the potential and the mass of the scalar degree of freedom. The slow-roll parameters $\epsilon$, $\eta$ and the $e$-fold number of the model were evaluated. The model gives $e$-fold number $N_e\approx 9.7$, characterizing the age of inflation, in the Einstein frame. We note that the values of the slow-roll parameters and $e$-foldings are different in the Einstein and Jordan frames. We show by the analysis of critical points of autonomous equations that the standard matter era exists and the standard matter era conditions are satisfied. We have obtained the approximate solution of equations of motion which is the deviation from the de Sitter phase. To verify that the model can be consistent with the accelerating expansion of the present universe if the curvature is small, one should solve the system of equation (27), (28). We leave such investigation for the future. It should be noted that the effective Newton constant in $F(R)$ gravity models is given by $G_{eff}=G/F'(R)$ and in the model (1) becomes $G_{eff}=G\sqrt{1-(\beta R)^2}$. Thus, in the small curvature regime, $\beta R\ll 1$, the corrections to the Newton law are negligible. It was shown that the model passes the matter stability test. The model may be alternative to GR, and can describe early-time inflation. The possible future singularities in this model were not investigated that we leave for further study.

\end{document}